\documentclass[twocolumn,showpacs,amsmath,amssymb,amsfonts,floatfix,prl,aps]{revtex4}
\usepackage{graphicx}
\usepackage{color}

\begin{document}
\title{Laser induced ultrafast demagnetization: an \emph{ab-initio} perspective}
\author{K. Krieger$^{1}$}
\author{J. K. Dewhurst$^{1}$}
\author{P. Elliott$^{1}$}
\author{S. Sharma$^{1,2}$}
\email{sharma@mpi-halle.mpg.de}
\author{E. K. U. Gross$^{1}$}
\affiliation{1. Max-Planck-Institut f\"ur Mikrostrukturphysik, Weinberg 2, 
D-06120 Halle, Germany.}
\affiliation{2. Department of physics, Indian Institute for technology-Roorkee, 247667 Uttarkhand, India}
\date{\today}

\begin{abstract}
Time-dependent density functional theory is implemented in an all electron solid-state code for the case of fully non-collinear spins. 
We use this to study laser induced demagnetization in Fe, Co and Ni. It is shown that this
demagnetization is a two-step process: excitation of a fraction of electrons followed by spin-flip transitions of the remaining localized
electrons. These results successfully explain several experimental features such as the time-lag between the start of the pulse and demagnetization and
spin-flip excitations dominating the physics. We further show that it is possible to control the moment loss by
tunable laser parameters like frequency, duration and intensity. 
\end{abstract}

\maketitle

Manipulation of electrons by femtosecond (fs) laser pulses opens up the vast and largely unexplored physical landscape of ultra-short time scales. 
One possibility in this landscape is to use electronic spins, which can be optically 
manipulated (flipped) using lasers to store data as binary bits. The advantage of such a technique would be a increase in the speed of data storage by orders of magnitude. Ultrafast light induced  demagnetization\cite{agranat84} was demonstrated in the 90s\cite{beaurepaire96}, 
where demagnetization times (in Ni) faster than a few pico-seconds were 
achieved using intense laser pulses\cite{vaterlaus92}. Recently, these demagnetization times have been measured down to a few hundred femtoseconds, owing to  
advances made in the refinement of pump-probe and other 
experiments\cite{beaurepaire96,hohlfeld97,scholl97,aeschlimann97,hohlfeld99,regensburger00,schmidt05,kirilyuk10,melnikov11}. 
However, we are still far from achieving sufficiently controlled 
manipulation\cite{kirilyuk10} of spins required for the production of useful devices. One of the reasons behind this is the lack of full 
understanding of the phenomena leading to laser induced demagnetization. 

There have been a number of attempts at explaining this laser induced loss of moment:
Combined action of spin-orbit coupling and interaction between spins and laser photons\cite{zhang00}.
Super-diffusive spin transport where excited electrons carry spin with them from one part of the sample to another\cite{battiato10,rhie03}.
Elliott-Yafet mechanism where electron-phonon or electron-impurity mediated spin-flip is the major contributor\cite{koopmans05,koopmans07}. 
Transfer of spin angular momentum to the lattice\cite{hubner96}. All these studies have in common that they 
describe the dynamics of the \emph{excited electrons} using parameterized model Hamiltonians\cite{popova11}. 

Time-dependent density functional theory (TDDFT)\cite{runge84},
which extends density functional theory into the time domain, is a formally exact method for describing the real-time 
dynamics of electrons under the influence of an external field such as the vector potential of the applied laser pulse.
The advantage of such a technique is clear: it does not require any empirical parameter, is fully \emph{ab-initio} and not
only linear, but also highly non-linear processes are a natural part of the simulation.

In the present work we use spin-resolved TDDFT to study the process of laser induced
demagnetization. Magnetic non-collinearity can be a major contributor in the loss of moment so in order not to exclude such effects we 
extended TDDFT to the fully non-collinear case. We have further implemented this fully non-collinear magnetic 
time propagation for periodic systems in an all-electron code. 
Several bulk systems (Fe, Co and Ni) are studied using this code in order to explore various possible demagnetization scenarios.
With optimal control of spins in mind we have also explored the effect 
of various tunable laser parameters of a laser pulse on the process of demagnetization.  

Our analysis shows that 
the demagnetization occurs as a two-step process, where first the electrons make transitions to excited states followed by 
spin-orbit mediated spin-flip transitions. The interesting point about these results is that
it is \emph{not the excited delocalized electrons} which make these spin-flip transitions leading to the demagnetization of the material,
but rather the remaining localized electrons. 
We note that while the non-collinearity of the spins is essential, long range coherent process like magnons do not play a significant 
role in the demagnetization in bulk systems for the time scales studied in the present work.
These results explain several experimental indications like time-lag between the start of the pulse and moment 
loss\cite{beaurepaire96,scholl97,sultan12} as well as
spin-flip excitations dominating the initial demagnetization process\cite{scholl97,rhie03}.
It is further demonstrated that the spins can be controlled using easily tunable laser parameters, for example
the magnitude of the loss in moment (in Ni) can be controlled by changing intensity, frequency and/or duration of the laser pulse.


The Runge-Gross theorem\cite{runge84} establishes that the time-dependent external potential is a unique functional of
the time-dependent density, given the initial state. 
Based on this theorem, a system of non-interacting particles can be chosen such that the density of this non-interacting system is 
equal to that of the interacting system \emph{for all times}. The wave function of this 
non-interacting system is represented as a Slater determinant of single-particle orbitals. In what follows we shall employ the 
non-collinear spin-dependent version of these theorems. Then the time-dependent Kohn-Sham (KS) orbitals are Pauli spinors determined 
by the equations:
\begin{eqnarray}
\label{hamil}
&i&\frac{\partial \psi_j({\bf r},t)}{\partial t}=\left[
\frac{1}{2}\left(-i{\mathbf \nabla} +\frac{1}{c}{\bf A}_{\rm ext}(t)\right)^2 +v_{s}({\bf r},t) \right. \\ \nonumber
&+&\left. 
\frac{1}{2c} \overrightarrow{\mathbf \sigma}\cdot{\bf B}_{s}({\bf r},t) +
\frac{1}{4c^2} \overrightarrow{\mathbf \sigma}\cdot ({\mathbf \nabla}v_{s}({\bf r},t) \times i{\mathbf \nabla})\right]
\psi_j({\bf r},t)
\end{eqnarray}
where ${\bf A}_{\rm ext}(t)$ is a vector potential representing the applied laser field, and ${\bf \sigma}$ are the Pauli matrices. 
The KS effective potential $v_{s}({\bf r},t) = v_{\rm ext}({\bf r},t)+v_{\rm H}({\bf r},t)+v_{\rm xc}({\bf r},t)$ is decomposed into the 
external potential $v_{\rm ext}$, the classical electrostatic Hartree potential $v_{\rm H}$ and the exchange-correlation potential $v_{\rm xc}$. 
Similarly the KS magnetic field is written as ${\bf B}_{s}({\bf r},t)={\bf B}_{\rm ext}(t)+{\bf B}_{\rm xc}({\bf r},t)$ where ${\bf B}_{\rm ext}(t)$ is 
the magnetic field of the applied laser pulse plus possibly an additional magnetic field and ${\bf B}_{\rm xc}({\bf r},t)$ is the XC magnetic field. The final term of Eq. (\ref{hamil}) is 
the spin-orbit coupling term. 
Since the wavelength of the applied laser in the present work is much greater than the size of a unit cell we apply the dipole approximation 
and hence disregard the spatial dependence of the vector potential.

The exchange-correlation potential has a functional dependence on the density and the magnetization density of the system at the current and 
all previous times. Hence it includes the information about the whole history of this time propagation. 
Knowledge of this functional would solve all time-dependent (externally driven) interacting problems. In practice 
however, the exchange-correlation potential is always approximated. 
In the present work we use the adiabatic local spin density approximation (ALSDA)\cite{Zangwill1980}. 
Using the method outlined above, various extended magnetic systems are studied\cite{param}
using the full-potential linearized augmented plane wave (FP-LAPW)
method \cite{Singh}, implemented within the Elk code \cite{elk}.

\begin{figure}[ht]
\centerline{\includegraphics[width=\columnwidth,angle=-0]{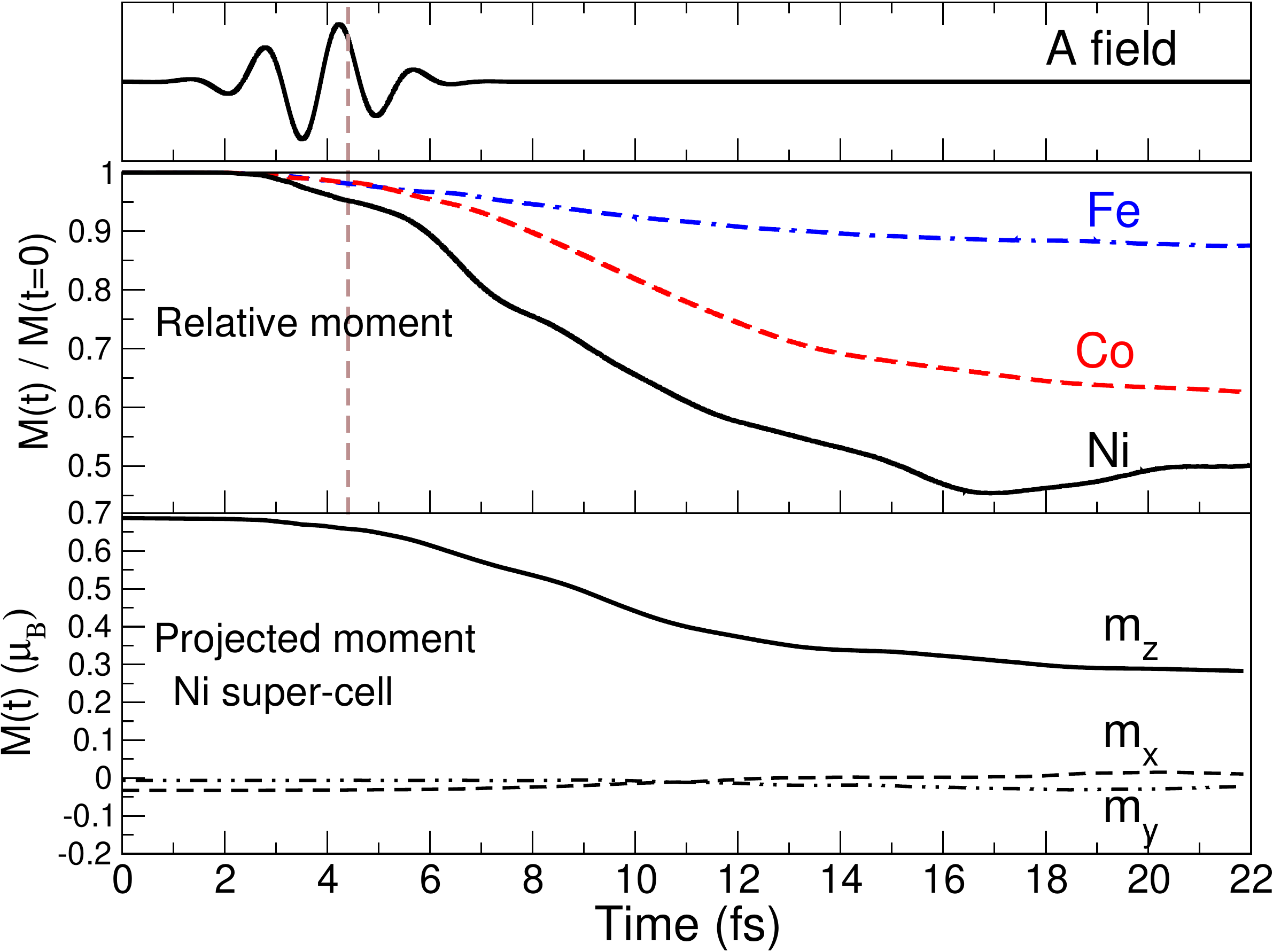}}
\caption{(Color online) Top panel: {\bf A}(t) of the laser pulse\cite{pulse1}. 
Middle panel: relative magnetic moment for, Fe, Co and Ni. Lower panel: $x$, $y$ and $z$-projected magnetic 
moment per atom (in Bohr magneton) for a super-cell of Ni. All times are in femtoseconds}
\label{feconi}
\end{figure}

Presented in the middle panel of Fig. \ref{feconi} are the magnetic moments of bulk Fe, Co and Ni as a function of time under the influence of an intense
laser pulse\cite{pulse1}. In all cases, demagnetization is observed-- the largest loss of moment is for
Ni (43\%) and the smallest for Fe (12\%). 
We observe that in all three cases the systems do not become non-collinear in the sense that the
loss of moment in the $z$-direction would be gained in the $x$ or $y$-direction. It may be argued that since these calculations are 
performed using  a single atom unit cell with periodic boundary conditions, it is premature to make any conclusions about the
contribution of non-collinearity to the loss of moment. Hence we have studied the effect of the same laser pulse on a Ni unit-cell 4 times as large in size.
So as not to bias our calculations towards collinearity we started (at 0 fs) from a random configuration of spins
with respect to one other in this super-cell and the results for the moment (per Ni atom) projected in $x$, $y$ and $z$-directions are presented in the
lower panel of Fig. \ref{feconi}.  
While it is essential that non-collinearity is included in the calculation (due to the presence of the spin-orbit coupling term in Eq. (\ref{hamil})), 
we find that long-range non-collinearity, like the relative alignment
of moments between atomic sites, does not play a significant role. 
 This is mostly due to the fact that for the small time scales of interest in the present work one does not expect 
low energy non-collinear processes like magnons or spin-waves to dominate. 
Interestingly  we note that the dimensionality plays an important role in this:
for lower-dimensional systems like mono-layers and super-structures, inter-site non-collinearity of spins was found to be important.

\begin{figure}[ht]
\centerline{\includegraphics[width=\columnwidth,angle=-0]{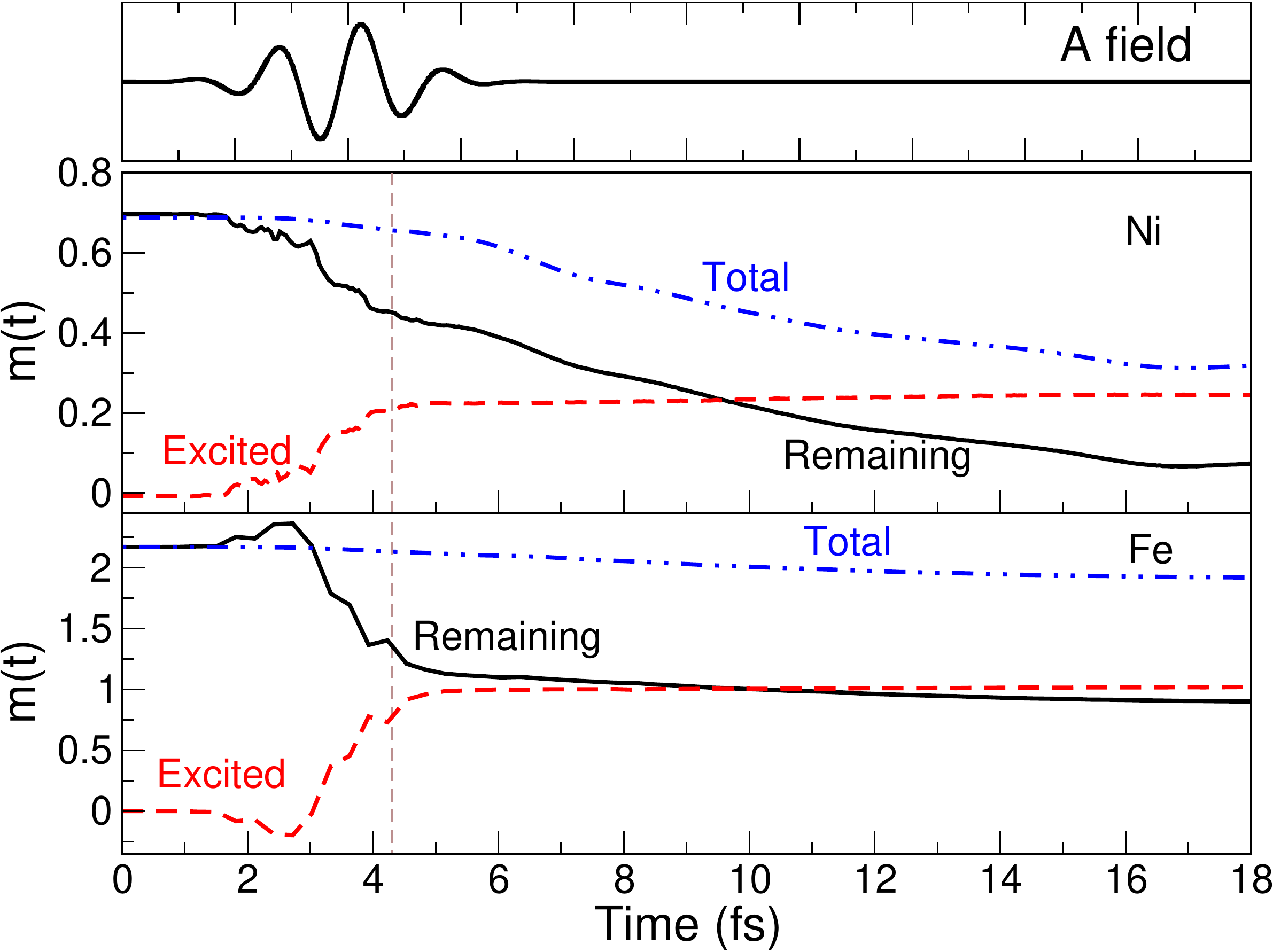}}
\caption{(Color online) Top panel: {\bf A}(t) of the laser pulse\cite{pulse1}. 
Middle panel: total magnetic moment per Ni atom, magnetic moment per Ni atom coming from excited and remaining electrons. 
Lower panel: same as middle panel but for Fe. All moments are in Bohr magneton and times in femtoseconds.}
\label{exnex}
\end{figure}
A feature of the demagnetization process in all three cases (Fe, Co and Ni) is a time lag of  $\sim$5 fs between the start of 
the laser pulse and the beginning of the loss in the magnetic moment. Such a time lag has been noted in almost all experiments with the actual value
of the time-lag depending upon the laser pulse used\cite{beaurepaire96,scholl97,guidoni02,zhang09,sultan12}. 
The questions now arise what is the origin of this time lag and what causes the demagnetization? 
To understand this, we plot in Fig. \ref{exnex} two contributions to the moment for Fe and Ni. They are 
(a) from the electrons which under the influence 
of intense laser pulse make a transition to excited states and become delocalized and 
(b) from the remaining electrons which are localized
close to the nuclei. During the first $\sim$5 fs the electrons, carrying their spins, make transitions to excited states. This leads to an increase in 
the total moment associated with the excited electrons and to a lowering of the moment coming from the remaining localized electrons such 
that the sum of the two moments stays almost constant. 
After $\sim$5 fs the moment of the excited delocalized electrons stays almost constant while some of the remaining localized electrons make spin-flip 
transitions leading to a loss in the total moment. Hence the demagnetization mechanism is clearly a two step process:
(1) during the first $\sim$5 fs a fraction of the electrons becomes delocalized by making transitions to the excited states 
(2) this is followed by the remaining localized electrons making spin-flip transitions. 
The major factor responsible for these spin-flip transitions of the localized electrons is the spin-orbit coupling\cite{stamm10,popova13} 
term in Eq. (\ref{hamil}). To confirm this fact we also performed similar calculations in which the spin-orbit coupling term was set to zero, and find no such demagnetization. 

This two-step process for demagnetization provides a clear explanation of the time-lag and is in line
with experimental work which indicates fast charge dynamics followed by slower spin-dynamics\cite{beaurepaire96,guidoni02}, and also with experiments
which show a difference in the dynamics of spins and orbital angular moment\cite{boeglin10}.
These results are also consistent with 
the idea that the indirect effect of spin-orbit coupling is slower than the direct absorption of light.   
These results also agree with the
experimental results which show that it is indeed the spin-flip excitations which lead to demagnetization in the first picosecond and the
non-collinear processes only start to have a substantial contribution much later\cite{scholl97,rhie03}.

In the present work periodic boundary conditions are used and the excited electrons can not move away (diffusive mechanisms like proposed 
in Refs. \cite{battiato10,melnikov11} are thus ignored). In a more realistic scenario such a diffusion of electrons will 
lead to further demagnetization. Such a process would also contribute in the absence of any spin-orbit coupling and 
we expect demagnetization to be more than what is seen here. 
Other phenomena like lattice vibrations or radiative effects are not expected to contribute substantially for the time scales 
relevant to the present work.

\begin{figure}[ht]
\centerline{\includegraphics[width=\columnwidth,angle=-0]{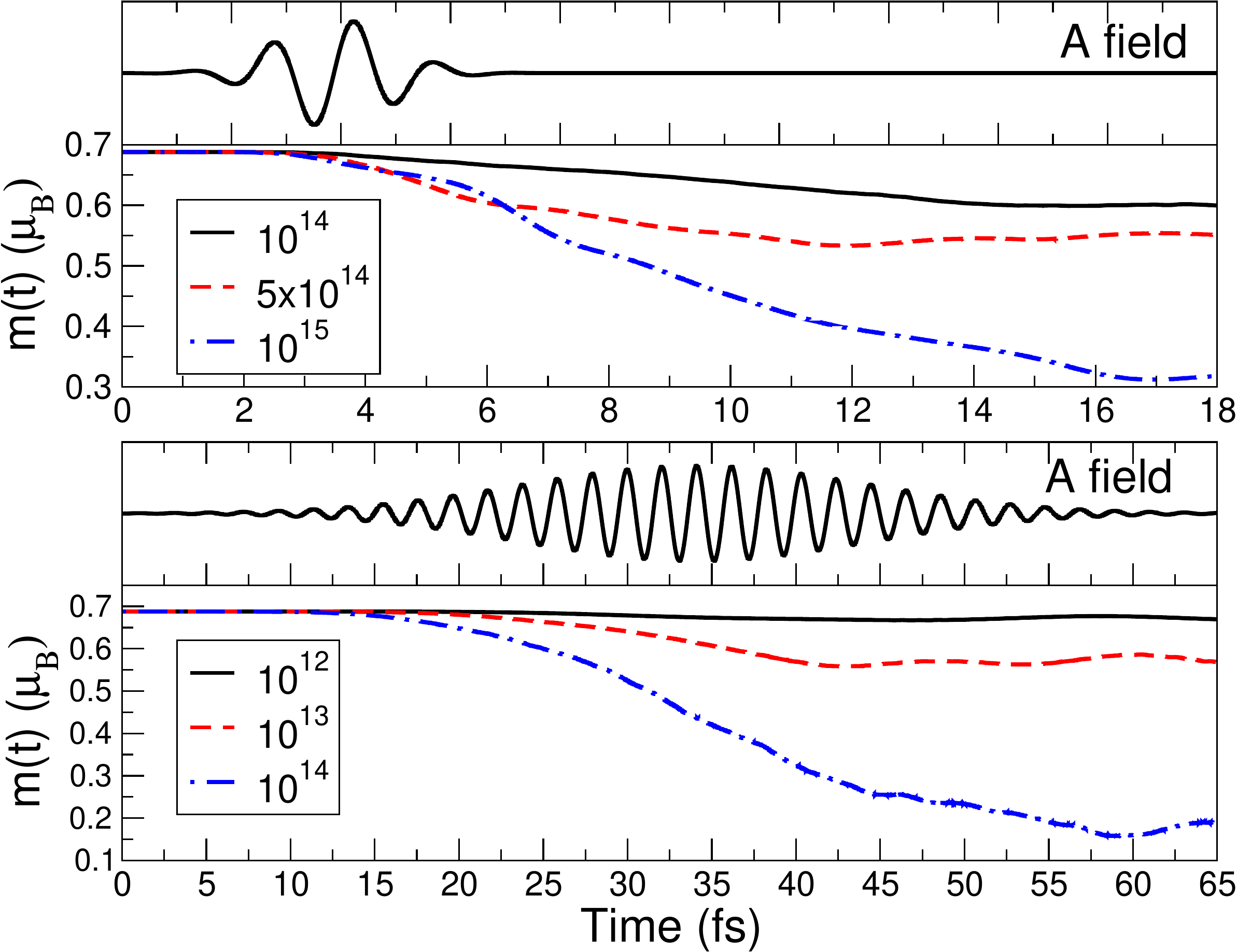}}
\caption{(Color online)Top panel: {\bf A}(t) of the laser pulse of duration 6 fs . 
Second panel: total magnetic moment per Ni atom under the influence of laser pulses 
with three different peak intensities (see Ref. \onlinecite{pulse2}). Third panel: {\bf A}(t) of the laser pulse of duration 60 fs. 
Lower panel: total magnetic moment per Ni atom under the influence of laser pulses with three different peak 
intensities (see Ref. \onlinecite{pulse3}). All moments are in Bohr magneton and times are in femtoseconds.}
\label{intens}
\end{figure}
The above analysis scrutinizes  the process of demagnetization. What is most important for future technological applications is not just this
knowledge but also the means to manipulate spins in a controlled manner. We now
show the effect of the three easily tunable parameters of a laser pulse; intensity, frequency and duration on demagnetization. In the upper panel of
Fig. \ref{intens} is
shown the magnetic moment for Ni as a function of the peak intensity\cite{pulse2}. The effect of intensity on the demagnetization 
is dramatic. For a pulse of peak intensity 10$^{15}$ W/cm$^2$ 35\% of the moment is lost after 20 fs while only
14\% loss is observed for a pulse of peak intensity 10$^{14}$ W/cm$^2$. It is important to mention that at these 
intensities the response of the system is far from linear and non-linear effects are predominant.

In all these studies, a very short laser pulse of 6 fs was applied\cite{pulse2}. 
Most experiments are currently limited to using much longer laser pulses. In the lower panel of Fig.  \ref{intens}, we show the effect of 
such a long pulse of varying intensities\cite{pulse3}. Again we find that demagnetization increases with increasing intensity. 
For longer duration pulses, a higher demagnetization, with 71\% loss in moment induced by the pulse of peak intensity 10$^{14}$ W/cm$^2$, 
is observed. Also a clear indication from Fig. \ref{intens} is that for longer duration pulses, a lower intensity is sufficient to obtain a large
demagnetization. We note that the time-lag between the start of the pulse and demagnetization depends upon the duration and frequency of 
the pulse and increases to $\sim$15 fs on going from a pulse of duration 6 fs to 60 fs.
Here we also observe that the physics of moment loss remains exactly the same i.e. a two-step 
demagnetization process with a time-lag. Hence the use of ultra short laser pulses in the present work is a matter of convenience for 
obtaining results within reasonable computer time by not having to propagate for very long times. An added benefit of short time scale pulses
is also that we can clearly separate the spin-dependent phenomena from charge only excitations, which gets mixed with each 
other on large time scales.

\begin{figure}[ht]
\centerline{\includegraphics[width=\columnwidth,angle=-0]{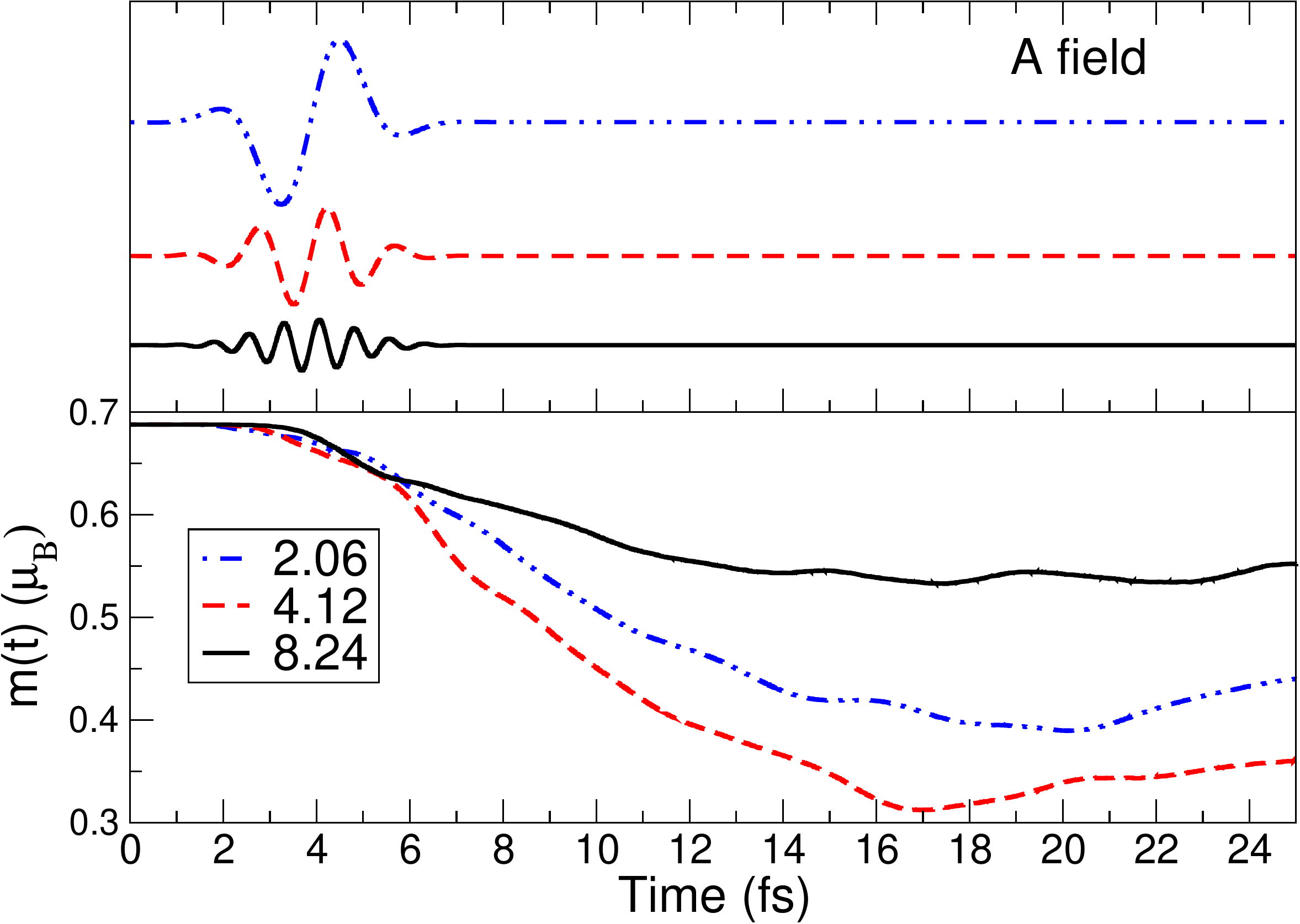}}
\caption{(Color online) Top panel: {\bf A}(t) of the laser pulses of peak intensity 10$^{15}$ W/cm$^2$ and varying
frequency\cite{pulse4}. Lower panel: total magnetic moment (in Bohr magneton) per Ni atom under the influence of 
laser pulses in the top panel. All times are in femtoseconds.}
\label{freq}
\end{figure}
Another laser beam parameter that affects the magnetic moment is the carrier frequency of the pulse. In Fig. \ref{freq} we present the results for 
short laser pulses (6 fs) of  peak intensity 10$^{15}$ W/cm$^2$ but with varying frequency.  It is clear that the central frequency of the pulse
can also be used to control the amount of demagnetization. The dependence of demagnetization on frequency is non-linear and  can be tuned to
obtain a loss in moment of between 20\% and 53\% for bulk Ni. The ideal frequency needed to achieve maximum moment loss (or rather at which the system
becomes most absorptive) is a material-dependent property and is related to the details of the band structure.

\begin{figure}[ht]
\centerline{\includegraphics[width=\columnwidth,angle=-0]{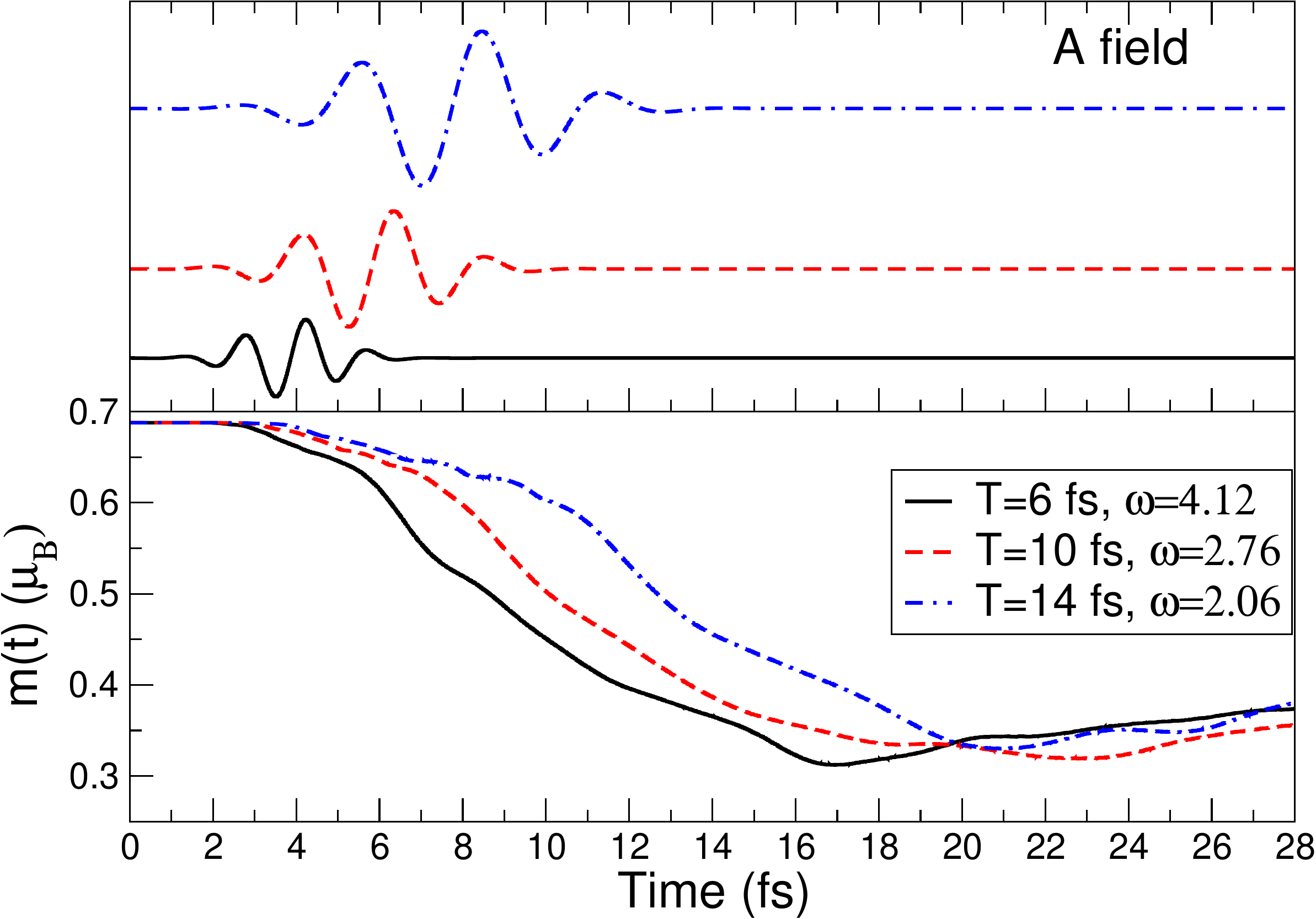}}
\caption{(Color online) Top panel: {\bf A}(t) of the laser pulses (see Ref. \onlinecite{pulse5}). 
Lower panel: total magnetic moment (in Bohr magneton) per Ni atom under the influence of laser pulses in the top panel. 
All times are in femtoseconds}
\label{cntrl}
\end{figure}
At this point we ask ourselves whether these three parameters can be jointly selected to get the desired demagnetization.
In Fig. \ref{cntrl} we show results for such a study. The peak intensity is kept fixed (10$^{15}$ W/cm$^2$) but the frequency and duration of the pulse are
tuned\cite{pulse5} to get the same degree of demagnetization at 20 fs. From these results we see that even though a particular set of frequency and duration
combinations follow totally different paths in magnetic phase space as a function of time, at the end of 20 fs the loss in moment is the same and remains 
almost the same after 20 fs. It is important to mention the effect of yet another laser parameter, namely the polarization of the pulse.
In the present work, linearly polarized light in $x$-direction was used (perpendicular to the direction of the moment, which points along the $z$-axis). 
Like experiments\cite{longa07}, we find that changing the plane of polarization of 
this linearly polarized light does not affect the process of demagnetization. In future it would be interesting to study the effect of circularly polarized light
on the process of demagnetization\cite{hansteen06}

To conclude: in the present work we show that by means of the first fully \emph{ab-initio} TDDFT study of Fe, Co and Ni,
laser induced demagnetization is a two-step process. Excitation of a fraction of electrons followed by spin-flip transitions of the remaining localized
electrons is a generic process appearing in all our calculations. These results can successfully explain several experimental features like time-lag between start of pulse
and demagnetization, spin-flip excitations dominating the physics initially and independence of loss of moment
from the polarization of the perturbing electromagnetic field.
Interestingly, we find that it is not the excited delocalized electrons which dominate the process of demagnetization
but rather the remaining localized electrons. 
Studying the effect of easily-tunable laser parameters on the process of moment loss in Ni 
provides insight into optimal control of the spin moment in solids. In particular, we note that the amount of demagnetization and the
time-lag between the start of the pulse and moment loss is strongly dependent on the intensity, frequency and duration of the perturbing laser pulse.


\begin{thebibliography}{36}
\expandafter\ifx\csname natexlab\endcsname\relax\def\natexlab#1{#1}\fi
\expandafter\ifx\csname bibnamefont\endcsname\relax
  \def\bibnamefont#1{#1}\fi
\expandafter\ifx\csname bibfnamefont\endcsname\relax
  \def\bibfnamefont#1{#1}\fi
\expandafter\ifx\csname citenamefont\endcsname\relax
  \def\citenamefont#1{#1}\fi
\expandafter\ifx\csname url\endcsname\relax
  \def\url#1{\texttt{#1}}\fi
\expandafter\ifx\csname urlprefix\endcsname\relax\def\urlprefix{URL }\fi
\providecommand{\bibinfo}[2]{#2}
\providecommand{\eprint}[2][]{\url{#2}}

\bibitem[{\citenamefont{Agranat et~al.}(1984)\citenamefont{Agranat, Ashitkov,
  Granovskii, and Rukman}}]{agranat84}
\bibinfo{author}{\bibfnamefont{M.~B.} \bibnamefont{Agranat}},
  \bibinfo{author}{\bibfnamefont{S.~I.} \bibnamefont{Ashitkov}},
  \bibinfo{author}{\bibfnamefont{A.~B.} \bibnamefont{Granovskii}},
  \bibnamefont{and} \bibinfo{author}{\bibfnamefont{G.~I.}
  \bibnamefont{Rukman}}, \bibinfo{journal}{Sov. Phys. JEPT}
  \textbf{\bibinfo{volume}{59}}, \bibinfo{pages}{804} (\bibinfo{year}{1984}).

\bibitem[{\citenamefont{Beaurepaire et~al.}(1996)\citenamefont{Beaurepaire,
  Merle, Daunois, and Bigot}}]{beaurepaire96}
\bibinfo{author}{\bibfnamefont{E.}~\bibnamefont{Beaurepaire}},
  \bibinfo{author}{\bibfnamefont{J.-C.} \bibnamefont{Merle}},
  \bibinfo{author}{\bibfnamefont{A.}~\bibnamefont{Daunois}}, \bibnamefont{and}
  \bibinfo{author}{\bibfnamefont{J.-Y.} \bibnamefont{Bigot}},
  \bibinfo{journal}{Phys. Rev. Lett.} \textbf{\bibinfo{volume}{76}},
  \bibinfo{pages}{4250} (\bibinfo{year}{1996}).

\bibitem[{\citenamefont{Vaterlaus et~al.}(1992)\citenamefont{Vaterlaus,
  Beutler, Guarisco, Lutz, and Meier}}]{vaterlaus92}
\bibinfo{author}{\bibfnamefont{A.}~\bibnamefont{Vaterlaus}},
  \bibinfo{author}{\bibfnamefont{T.}~\bibnamefont{Beutler}},
  \bibinfo{author}{\bibfnamefont{D.}~\bibnamefont{Guarisco}},
  \bibinfo{author}{\bibfnamefont{M.}~\bibnamefont{Lutz}}, \bibnamefont{and}
  \bibinfo{author}{\bibfnamefont{F.}~\bibnamefont{Meier}},
  \bibinfo{journal}{Phys. Rev. B} \textbf{\bibinfo{volume}{46}},
  \bibinfo{pages}{5280} (\bibinfo{year}{1992}).

\bibitem[{\citenamefont{Hohlfeld et~al.}(1997)\citenamefont{Hohlfeld, Matthias,
  Knorren, and Bennemann}}]{hohlfeld97}
\bibinfo{author}{\bibfnamefont{J.}~\bibnamefont{Hohlfeld}},
  \bibinfo{author}{\bibfnamefont{E.}~\bibnamefont{Matthias}},
  \bibinfo{author}{\bibfnamefont{R.}~\bibnamefont{Knorren}}, \bibnamefont{and}
  \bibinfo{author}{\bibfnamefont{K.~H.} \bibnamefont{Bennemann}},
  \bibinfo{journal}{Phys. Rev. Lett.} \textbf{\bibinfo{volume}{78}},
  \bibinfo{pages}{4861} (\bibinfo{year}{1997}).

\bibitem[{\citenamefont{Scholl et~al.}(1997)\citenamefont{Scholl, Baumgarten,
  Jacquemin, and Eberhardt}}]{scholl97}
\bibinfo{author}{\bibfnamefont{A.}~\bibnamefont{Scholl}},
  \bibinfo{author}{\bibfnamefont{L.}~\bibnamefont{Baumgarten}},
  \bibinfo{author}{\bibfnamefont{R.}~\bibnamefont{Jacquemin}},
  \bibnamefont{and}
  \bibinfo{author}{\bibfnamefont{W.}~\bibnamefont{Eberhardt}},
  \bibinfo{journal}{Phys. Rev. Lett.} \textbf{\bibinfo{volume}{79}},
  \bibinfo{pages}{5146} (\bibinfo{year}{1997}).

\bibitem[{\citenamefont{Aeschlimann et~al.}(1997)\citenamefont{Aeschlimann,
  Bauer, Pawlik, Weber, Burgermeister, Oberli, and Siegmann}}]{aeschlimann97}
\bibinfo{author}{\bibfnamefont{M.}~\bibnamefont{Aeschlimann}},
  \bibinfo{author}{\bibfnamefont{M.}~\bibnamefont{Bauer}},
  \bibinfo{author}{\bibfnamefont{S.}~\bibnamefont{Pawlik}},
  \bibinfo{author}{\bibfnamefont{W.}~\bibnamefont{Weber}},
  \bibinfo{author}{\bibfnamefont{R.}~\bibnamefont{Burgermeister}},
  \bibinfo{author}{\bibfnamefont{D.}~\bibnamefont{Oberli}}, \bibnamefont{and}
  \bibinfo{author}{\bibfnamefont{H.~C.} \bibnamefont{Siegmann}},
  \bibinfo{journal}{Phys. Rev. Lett.} \textbf{\bibinfo{volume}{79}},
  \bibinfo{pages}{5158} (\bibinfo{year}{1997}).

\bibitem[{\citenamefont{Hohlfeld et~al.}(1999)\citenamefont{Hohlfeld, G\"udde,
  D\"uhr, Korn, and Matthias}}]{hohlfeld99}
\bibinfo{author}{\bibfnamefont{J.}~\bibnamefont{Hohlfeld}},
  \bibinfo{author}{\bibfnamefont{J.}~\bibnamefont{G\"udde}},
  \bibinfo{author}{\bibfnamefont{U.~C.~O.} \bibnamefont{D\"uhr}},
  \bibinfo{author}{\bibfnamefont{G.}~\bibnamefont{Korn}}, \bibnamefont{and}
  \bibinfo{author}{\bibfnamefont{E.}~\bibnamefont{Matthias}},
  \bibinfo{journal}{Appl. Phys. B: Lasers Opt.} \textbf{\bibinfo{volume}{68}},
  \bibinfo{pages}{505} (\bibinfo{year}{1999}).

\bibitem[{\citenamefont{Regensburger et~al.}(2000)\citenamefont{Regensburger,
  Vollmer, and Kirschner}}]{regensburger00}
\bibinfo{author}{\bibfnamefont{H.}~\bibnamefont{Regensburger}},
  \bibinfo{author}{\bibfnamefont{R.}~\bibnamefont{Vollmer}}, \bibnamefont{and}
  \bibinfo{author}{\bibfnamefont{J.}~\bibnamefont{Kirschner}},
  \bibinfo{journal}{Phys. Rev. B} \textbf{\bibinfo{volume}{61}},
  \bibinfo{pages}{14716} (\bibinfo{year}{2000}).

\bibitem[{\citenamefont{Schmidt et~al.}(2005)\citenamefont{Schmidt, Pickel,
  Wiemh\"ofer, Donath, and Weinelt}}]{schmidt05}
\bibinfo{author}{\bibfnamefont{A.~B.} \bibnamefont{Schmidt}},
  \bibinfo{author}{\bibfnamefont{M.}~\bibnamefont{Pickel}},
  \bibinfo{author}{\bibfnamefont{M.}~\bibnamefont{Wiemh\"ofer}},
  \bibinfo{author}{\bibfnamefont{M.}~\bibnamefont{Donath}}, \bibnamefont{and}
  \bibinfo{author}{\bibfnamefont{M.}~\bibnamefont{Weinelt}},
  \bibinfo{journal}{Phys. Rev. Lett.} \textbf{\bibinfo{volume}{95}},
  \bibinfo{pages}{107402} (\bibinfo{year}{2005}).

\bibitem[{\citenamefont{Kirilyuk et~al.}(2010)\citenamefont{Kirilyuk, Kimel,
  and Rasing}}]{kirilyuk10}
\bibinfo{author}{\bibfnamefont{A.}~\bibnamefont{Kirilyuk}},
  \bibinfo{author}{\bibfnamefont{A.~V.} \bibnamefont{Kimel}}, \bibnamefont{and}
  \bibinfo{author}{\bibfnamefont{T.}~\bibnamefont{Rasing}},
  \bibinfo{journal}{Rev. Mod. Phys.} \textbf{\bibinfo{volume}{82}},
  \bibinfo{pages}{2731} (\bibinfo{year}{2010}).

\bibitem[{\citenamefont{Melnikov et~al.}(2011)\citenamefont{Melnikov,
  Razdolski, Wehling, Papaioannou, Roddatis, Fumagalli, Aktsipetrov,
  Lichtenstein, and Bovensiepen}}]{melnikov11}
\bibinfo{author}{\bibfnamefont{A.}~\bibnamefont{Melnikov}},
  \bibinfo{author}{\bibfnamefont{I.}~\bibnamefont{Razdolski}},
  \bibinfo{author}{\bibfnamefont{T.~O.} \bibnamefont{Wehling}},
  \bibinfo{author}{\bibfnamefont{T.~E.} \bibnamefont{Papaioannou}},
  \bibinfo{author}{\bibfnamefont{V.}~\bibnamefont{Roddatis}},
  \bibinfo{author}{\bibfnamefont{P.}~\bibnamefont{Fumagalli}},
  \bibinfo{author}{\bibfnamefont{O.}~\bibnamefont{Aktsipetrov}},
  \bibinfo{author}{\bibfnamefont{A.~I.} \bibnamefont{Lichtenstein}},
  \bibnamefont{and}
  \bibinfo{author}{\bibfnamefont{U.}~\bibnamefont{Bovensiepen}},
  \bibinfo{journal}{Phys. Rev. Lett.} \textbf{\bibinfo{volume}{107}},
  \bibinfo{pages}{076601} (\bibinfo{year}{2011}).

\bibitem[{\citenamefont{Zhang and H\"ubner}(2000)}]{zhang00}
\bibinfo{author}{\bibfnamefont{G.~P.} \bibnamefont{Zhang}} \bibnamefont{and}
  \bibinfo{author}{\bibfnamefont{W.}~\bibnamefont{H\"ubner}},
  \bibinfo{journal}{Phys. Rev. Lett.} \textbf{\bibinfo{volume}{85}},
  \bibinfo{pages}{3025} (\bibinfo{year}{2000}).

\bibitem[{\citenamefont{Battiato et~al.}(2010)\citenamefont{Battiato, Carva,
  and Oppeneer}}]{battiato10}
\bibinfo{author}{\bibfnamefont{M.}~\bibnamefont{Battiato}},
  \bibinfo{author}{\bibfnamefont{K.}~\bibnamefont{Carva}}, \bibnamefont{and}
  \bibinfo{author}{\bibfnamefont{P.~M.} \bibnamefont{Oppeneer}},
  \bibinfo{journal}{Phys. Rev. Lett.} \textbf{\bibinfo{volume}{105}},
  \bibinfo{pages}{027203} (\bibinfo{year}{2010}).

\bibitem[{\citenamefont{Rhie et~al.}(2003)\citenamefont{Rhie, D\"urr, and
  Eberhardt}}]{rhie03}
\bibinfo{author}{\bibfnamefont{H.~S.} \bibnamefont{Rhie}},
  \bibinfo{author}{\bibfnamefont{H.~A.} \bibnamefont{D\"urr}},
  \bibnamefont{and}
  \bibinfo{author}{\bibfnamefont{W.}~\bibnamefont{Eberhardt}},
  \bibinfo{journal}{Phys. Rev. Lett.} \textbf{\bibinfo{volume}{90}},
  \bibinfo{pages}{247201} (\bibinfo{year}{2003}).

\bibitem[{\citenamefont{Koopmans et~al.}(2005)\citenamefont{Koopmans, Ruigrok,
  Dalla, and de~Jonge}}]{koopmans05}
\bibinfo{author}{\bibfnamefont{B.}~\bibnamefont{Koopmans}},
  \bibinfo{author}{\bibfnamefont{J.~J.~M.} \bibnamefont{Ruigrok}},
  \bibinfo{author}{\bibfnamefont{F.~L.} \bibnamefont{Dalla}}, \bibnamefont{and}
  \bibinfo{author}{\bibfnamefont{W.~J.~M.} \bibnamefont{de~Jonge}},
  \bibinfo{journal}{Phys. Rev. Lett.} \textbf{\bibinfo{volume}{95}},
  \bibinfo{pages}{267207} (\bibinfo{year}{2005}).

\bibitem[{\citenamefont{Koopmans}(2007)}]{koopmans07}
\bibinfo{author}{\bibfnamefont{B.}~\bibnamefont{Koopmans}},
  \bibinfo{journal}{Nature Mater.} \textbf{\bibinfo{volume}{6}},
  \bibinfo{pages}{715} (\bibinfo{year}{2007}).

\bibitem[{\citenamefont{H\"ubner and Bennemann}(1996)}]{hubner96}
\bibinfo{author}{\bibfnamefont{W.}~\bibnamefont{H\"ubner}} \bibnamefont{and}
  \bibinfo{author}{\bibfnamefont{K.~H.} \bibnamefont{Bennemann}},
  \bibinfo{journal}{Phys. Rev. B} \textbf{\bibinfo{volume}{53}},
  \bibinfo{pages}{3422} (\bibinfo{year}{1996}).

\bibitem[{\citenamefont{Popova et~al.}(2011)\citenamefont{Popova, Bringer, and
  Bl\"ugel}}]{popova11}
\bibinfo{author}{\bibfnamefont{D.}~\bibnamefont{Popova}},
  \bibinfo{author}{\bibfnamefont{A.}~\bibnamefont{Bringer}}, \bibnamefont{and}
  \bibinfo{author}{\bibfnamefont{S.}~\bibnamefont{Bl\"ugel}},
  \bibinfo{journal}{Phys. Rev. B} \textbf{\bibinfo{volume}{84}},
  \bibinfo{pages}{214421} (\bibinfo{year}{2011}).

\bibitem[{\citenamefont{Runge and Gross}(1984)}]{runge84}
\bibinfo{author}{\bibfnamefont{E.}~\bibnamefont{Runge}} \bibnamefont{and}
  \bibinfo{author}{\bibfnamefont{E.~K.~U.} \bibnamefont{Gross}},
  \bibinfo{journal}{Phys. Rev. Lett.} \textbf{\bibinfo{volume}{52}},
  \bibinfo{pages}{997} (\bibinfo{year}{1984}).

\bibitem[{\citenamefont{Sultan et~al.}(2012)\citenamefont{Sultan, Atxitia,
  Melnikov, Chubykalo-Fesenko, and Bovensiepen}}]{sultan12}
\bibinfo{author}{\bibfnamefont{M.}~\bibnamefont{Sultan}},
  \bibinfo{author}{\bibfnamefont{U.}~\bibnamefont{Atxitia}},
  \bibinfo{author}{\bibfnamefont{A.}~\bibnamefont{Melnikov}},
  \bibinfo{author}{\bibfnamefont{O.}~\bibnamefont{Chubykalo-Fesenko}},
  \bibnamefont{and}
  \bibinfo{author}{\bibfnamefont{U.}~\bibnamefont{Bovensiepen}},
  \bibinfo{journal}{Phys. Rev. B} \textbf{\bibinfo{volume}{85}},
  \bibinfo{pages}{184407} (\bibinfo{year}{2012}).

\bibitem[{\citenamefont{Zangwill and Soven}(1980)}]{Zangwill1980}
\bibinfo{author}{\bibfnamefont{A.}~\bibnamefont{Zangwill}} \bibnamefont{and}
  \bibinfo{author}{\bibfnamefont{P.}~\bibnamefont{Soven}},
  \bibinfo{journal}{Phys. Rev. Lett.} \textbf{\bibinfo{volume}{45}},
  \bibinfo{pages}{204} (\bibinfo{year}{1980}).

\bibitem[{par()}]{param}
\emph{\bibinfo{title}{\rm {In all cases a $k$-point mesh of $8\times8\times8$
  used and 120 empty states were needed for convergence. The symmetries were
  not used to reduce the $k$-point set. Time between 0-30 fs was divided into a
  total of 48000 time steps. Lattice parameter of 3.52 \AA\ for fcc Ni, 2.87
  \AA\ for bcc Fe and 3.544 \AA\ for fcc Co was used.}}}

\bibitem[{\citenamefont{Singh}(1994)}]{Singh}
\bibinfo{author}{\bibfnamefont{D.~J.} \bibnamefont{Singh}},
  \emph{\bibinfo{title}{\rm {Planewaves Pseudopotentials and the LAPW Method},
  {Kluwer Academic Publishers, Boston}}} (\bibinfo{year}{1994}).

\bibitem[{elk(2004)}]{elk}
 (\bibinfo{year}{2004}), \urlprefix\url{http://elk.sourceforge.net}.

\bibitem[{pul({\natexlab{a}})}]{pulse1}
\emph{\bibinfo{title}{\rm {The pulse duration (time between the start and the
  end of the pulse) is 6 fs, peak intensity is 10$^{15}$ W/cm$^2$, frequency is
  4.12/fs and the fluence is 934.8 mJ/cm$^2$. The pulse is linearly polarized
  along the $x$-axis perpendicular to the direction of the moment.}}}

\bibitem[{\citenamefont{Guidoni et~al.}(2002)\citenamefont{Guidoni,
  beaurepaire, and Bigot}}]{guidoni02}
\bibinfo{author}{\bibfnamefont{L.}~\bibnamefont{Guidoni}},
  \bibinfo{author}{\bibfnamefont{E.}~\bibnamefont{beaurepaire}},
  \bibnamefont{and} \bibinfo{author}{\bibfnamefont{J.~Y.} \bibnamefont{Bigot}},
  \bibinfo{journal}{Phys. Rev. Lett.} \textbf{\bibinfo{volume}{89}},
  \bibinfo{pages}{017401} (\bibinfo{year}{2002}).

\bibitem[{\citenamefont{Zhang et~al.}(2009)\citenamefont{Zhang, H\"ubner,
  Lefkidis, Bai, and George}}]{zhang09}
\bibinfo{author}{\bibfnamefont{G.~P.} \bibnamefont{Zhang}},
  \bibinfo{author}{\bibfnamefont{W.}~\bibnamefont{H\"ubner}},
  \bibinfo{author}{\bibfnamefont{G.}~\bibnamefont{Lefkidis}},
  \bibinfo{author}{\bibfnamefont{Y.}~\bibnamefont{Bai}}, \bibnamefont{and}
  \bibinfo{author}{\bibfnamefont{T.~F.} \bibnamefont{George}},
  \bibinfo{journal}{Nature Phys. Lett.} \textbf{\bibinfo{volume}{5}},
  \bibinfo{pages}{499} (\bibinfo{year}{2009}).

\bibitem[{\citenamefont{Stamm et~al.}(2010)\citenamefont{Stamm, Pontius,
  Kachel, Wietstruk, and D\"urr}}]{stamm10}
\bibinfo{author}{\bibfnamefont{C.}~\bibnamefont{Stamm}},
  \bibinfo{author}{\bibfnamefont{N.}~\bibnamefont{Pontius}},
  \bibinfo{author}{\bibfnamefont{T.}~\bibnamefont{Kachel}},
  \bibinfo{author}{\bibfnamefont{M.}~\bibnamefont{Wietstruk}},
  \bibnamefont{and} \bibinfo{author}{\bibfnamefont{H.~A.}
  \bibnamefont{D\"urr}}, \bibinfo{journal}{Phys. Rev. B}
  \textbf{\bibinfo{volume}{81}}, \bibinfo{pages}{104425}
  (\bibinfo{year}{2010}).

\bibitem[{\citenamefont{Popova et~al.}(2012)\citenamefont{Popova, Bringer, and
  Bl\"ugel}}]{popova13}
\bibinfo{author}{\bibfnamefont{D.}~\bibnamefont{Popova}},
  \bibinfo{author}{\bibfnamefont{A.}~\bibnamefont{Bringer}}, \bibnamefont{and}
  \bibinfo{author}{\bibfnamefont{S.}~\bibnamefont{Bl\"ugel}},
  \bibinfo{journal}{Phys. Rev. B} \textbf{\bibinfo{volume}{85}},
  \bibinfo{pages}{094419} (\bibinfo{year}{2012}).

\bibitem[{\citenamefont{Boeglin et~al.}(2010)\citenamefont{Boeglin,
  Beaurepaire, Halt\'e, L\'opez-Flores, Stamm, Pontius, D\"urr, and
  Bigot}}]{boeglin10}
\bibinfo{author}{\bibfnamefont{C.}~\bibnamefont{Boeglin}},
  \bibinfo{author}{\bibfnamefont{E.}~\bibnamefont{Beaurepaire}},
  \bibinfo{author}{\bibfnamefont{V.}~\bibnamefont{Halt\'e}},
  \bibinfo{author}{\bibfnamefont{V.}~\bibnamefont{L\'opez-Flores}},
  \bibinfo{author}{\bibfnamefont{C.}~\bibnamefont{Stamm}},
  \bibinfo{author}{\bibfnamefont{N.}~\bibnamefont{Pontius}},
  \bibinfo{author}{\bibfnamefont{H.~A.} \bibnamefont{D\"urr}},
  \bibnamefont{and} \bibinfo{author}{\bibfnamefont{J.-Y.} \bibnamefont{Bigot}},
  \bibinfo{journal}{Nature} \textbf{\bibinfo{volume}{465}},
  \bibinfo{pages}{458} (\bibinfo{year}{2010}).

\bibitem[{pul({\natexlab{b}})}]{pulse2}
\emph{\bibinfo{title}{\rm {The pulse duration (time between the start and the
  end of the pulse) is 6 fs and frequency is 4.12/fs. The pulse is linearly
  polarized along the $x$-axis perpendicular to the direction of the moment.
  The fluence is 934.8 mJ/cm$^2$, 467.5 mJ/cm$^2$ and 93.5 mJ/cm$^2$ for the
  pulse of peak intensity is 10$^{15}$ W/cm$^2$, 5$\times$10$^{14}$ W/cm$^2$
  and 10$^{14}$ W/cm$^2$ respectively.}}}

\bibitem[{pul({\natexlab{c}})}]{pulse3}
\emph{\bibinfo{title}{\rm {The pulse duration (time between the start and the
  end of the pulse) is 60 fs and frequency is 3.03/fs. The pulse is linearly
  polarized along the $x$-axis perpendicular to the direction of the moment.
  The fluence is 913.5 mJ/cm$^2$, 92.6 mJ/cm$^2$ and 9.3 mJ/cm$^2$ for the
  pulse of peak intensity is 10$^{14}$ W/cm$^2$, 10$^{13}$ W/cm$^2$ and
  10$^{12}$ W/cm$^2$ respectively.}}}

\bibitem[{pul({\natexlab{d}})}]{pulse4}
\emph{\bibinfo{title}{\rm {The pulse duration (time between the start and the
  end of the pulse) is 6 fs and the peak intensity is 10$^{15}$ W/cm$^2$. The
  pulse is linearly polarized along the $x$-axis perpendicular to the direction
  of the moment. The fluence is 1009.4 mJ/cm$^2$, 934.8 mJ/cm$^2$ and 915.9
  mJ/cm$^2$ for the pulse of frequency 2.06/fs, 4.14/fs and 8.24/fs
  respectively.}}}

\bibitem[{pul({\natexlab{e}})}]{pulse5}
\emph{\bibinfo{title}{\rm {The pulse is linearly polarized along the $x$-axis
  perpendicular to the direction of the moment and the peak intensity is
  10$^{15}$ W/cm$^2$. The fluence is 934.8 mJ/cm$^2$ for the pulse of duration
  6 fs and frequency 4.12/fs, 1402.2 mJ/cm$^2$ for the pulse of duration 10 fs
  and frequency 2.76/fs and 1869.6 mJ/cm$^2$ for the pulse of duration 14 fs
  and frequency 2.06/fs.}}}

\bibitem[{\citenamefont{Longa et~al.}(2007)\citenamefont{Longa, Kohlhepp,
  de~Jonge, and Koopmans}}]{longa07}
\bibinfo{author}{\bibfnamefont{F.~D.} \bibnamefont{Longa}},
  \bibinfo{author}{\bibfnamefont{J.~T.} \bibnamefont{Kohlhepp}},
  \bibinfo{author}{\bibfnamefont{W.~J.~M.} \bibnamefont{de~Jonge}},
  \bibnamefont{and} \bibinfo{author}{\bibfnamefont{B.}~\bibnamefont{Koopmans}},
  \bibinfo{journal}{Phys. Rev. B} \textbf{\bibinfo{volume}{75}},
  \bibinfo{pages}{224431} (\bibinfo{year}{2007}).

\bibitem[{\citenamefont{Hansteen et~al.}(2006)\citenamefont{Hansteen, Kimel,
  Kirilyuk, and Rasing}}]{hansteen06}
\bibinfo{author}{\bibfnamefont{F.}~\bibnamefont{Hansteen}},
  \bibinfo{author}{\bibfnamefont{A.}~\bibnamefont{Kimel}},
  \bibinfo{author}{\bibfnamefont{A.}~\bibnamefont{Kirilyuk}}, \bibnamefont{and}
  \bibinfo{author}{\bibfnamefont{T.}~\bibnamefont{Rasing}},
  \bibinfo{journal}{Phys. Rev. B} \textbf{\bibinfo{volume}{73}},
  \bibinfo{pages}{014421} (\bibinfo{year}{2006}).

\end{thebibliography}
\end{document}